_______________________________________________________________

# Dynamics of Explosive Events Observed by the Interface Region Imaging Spectrograph

E. Tavabi [1] · S. Zeighami [2*] · M. Heydari [1]

**Abstract** In this research, we investigate Explosive Events (EEs) in the off-limb solar atmosphere, with simultaneous observations from the Si IV, Mg II k, and slit-jaw images (SJI) of the Interface Region Imaging Spectrograph (IRIS), on 17 August 2014, and 19 February. IRIS data can be investigated to observe the motion of matter, fluctuations, energy absorption, and heat transition of the solar atmosphere. Mechanisms responsible for solar large-scale structures, such as flares and coronal mass ejections, might originate from these small-scale energetic events. Therefore, the study of these events can be helpful for understanding mechanisms in mass and energy transport from the chromosphere toward the transition region and corona. We obtain intensity profiles from spectra in two altitudes, i.e., at the solar limb and 5 arcsec distance from solar limb, and then analyze the EE fluctuations at these two altitudes along the slit. We find that some spectral line profiles show enhancements in blue and red wings indicating upward and downward flows, and some profiles have opposite EEs in both wings. The amplitude of the Doppler velocity in the two data sets of different altitudes was approximated to be about 50 km s$^{-1}$. We calculated the phase velocity of the oscillations using a technique based on cross-correlation. The phase velocity is obtained as about 220 km s$^{-1}$. According to the periodic red and blue enhancements in EEs, we suggest that the fluctuations in the EEs with one side enhancement indicate a swaying motions of spicules about their axes, and those EEs observed in both wings indicate a rotational motions of spicules. The swaying and rotational motions are indicative of kink and torsional waves, respectively.

_______________________________

✉ E. Tavabi
   e_tavabi@pnu.ac.ir

✉ S. Zeighami

zeighami@iaut.ac.ir

M. Heydari

mehriheidari95@yahoo.com

[1] Physics Department, Payame Noor University (PNU), 19395-3697-Tehran, I. R. of Iran

[2] Department of Physics, Tabriz Branch, Islamic Azad University, Tabriz, Iran

## 1. Introduction

EEs are dynamic events that first were observed in the solar transition region (TR) with the High-Resolution Telescope and Spectrograph (HRTS) (Brueckner and Bartoe, 1983). EEs could generate non-Gaussian profiles with strong enhancement. EEs have a spatial extension of $1 - 2$ arcsec, and a typical lifetime of roughly 60 s (Chen et al., 2019). It was first believed that such spectra are emitted from turbulent events (Brueckner and Bartoe, 1983), and for this reason, they are called explosive events (Dere, Bartoe, and Brueckner, 1984). Thereafter the term transition region explosive events (TREEs) is also used (Dere, Bartoe, and Brueckner, 1984). Some EEs could also produce enhancements in spectra of C I, C II, O I, and Mg II (Dere, Bartoe, and Brueckner, 1989; Dere, 1992; Zhang et al., 2010, and Tavabi and Koutchmy, 2019). EEs are repeatedly noted in observations (Chae et al., 1998, Ning, Innes, and Solanki, 2004, and Doyle, Popescu, and Taroyan, 2006). The were proposed to be caused by magnetic reconnection in the upper chromosphere modulated by p-mode oscillations. Non-Gaussian line profiles of EEs could be generated by spinning, unwinding, or twisting motions (Curdt, Tian, and Kamio, 2012, De Pontieu et al., 2014). In De Pontieu et al. (2014), the authors found that twisting motions of small-scale transition region (TR) loops and jets can result in EE-type line profiles of EEs. In Chen et al. (2019), using simultaneous imaging and spectroscopic observations from IRIS, the link between EEs and network jets was investigated. Their analysis suggests that some EEs are related to the birth or propagation of network jets, and the others are not connected to network jets. The energy source required to heat the solar corona plasma to a temperature of more than one million Kelvin in the dynamic solar photosphere is debated in solar physics. The transmission of energy through waves and oscillations can play an important role in understanding the dynamics of solar structures and the cause of the sudden rise of temperature in the solar atmosphere to several million Kelvin from the transition region to the solar corona (De Pontieu et al., 2007b). One of the energy transfer mechanisms is the propagation of magneto-hydrodynamic waves (Roberts, 2004). These waves in photospheric magnetic tubes can be generated by granular motions and then propagate along the chromosphere and penetrate the corona to transfer energy. Therefore, observations of oscillating motions in the solar atmosphere are a key test for the theory of coronal heating. The chromospheric layer of the solar atmosphere is composed of structures called spicules. The cut-off period in the chromosphere is 3 minutes (De Pontieu, Erdélyi, and James, 2004). This means that only waves with a period of less than 3 minutes can penetrate the higher layers. Spicules have diameters from about 120 km to 700 km, and their maximum lengths vary from a few hundred kilometers to 10 000 km, with most below 5000 km (De Pontieu et al., 2007b, 2012). Chromospheric spicules at the limb in coronal holes are intimately linked to the formation of features at TR and coronal temperatures (De Pontieu et al., 2007b). Jets are narrow plasma eruptions about 150 to 500 kilometers in diameter launched from the top of an almost homogeneous layer that stretches for about 3000 kilometers. Needleshaped spicules accumulate in areas of the solar supergranular network. Spicules are plasma explosions that occur almost every 5 minutes. Spicules are present even in the quiet Sun. Spicules are ubiquitous, high-velocity jets. Combining recent spectroscopic images of projectiles with solar observations has made it possible to study spicules' motion and their disappearance (Tavabi, Koutchmy, and Golub, 2015; Tavabi et al., 2015b). Recent studies have shown that spicules appear in several TR and chromospheric lines of the solar atmosphere. It is speculated that after disappearing from chromospheric lines, the spicules appear as jets in the transition region in hot regions (De Pontieu et al., 2007b and 2011, Pereira et al., 2014). The purpose of studying the spicules and their properties in different parts of the Sun is to obtain information about the chromosphere, which is a dynamic region and is of special importance in solar physics. As density is orders of magnitude lower in the corona than in the chromosphere, spicules play an important role in the mass equilibrium of the solar corona. The brightness of the spicules changes with temperature and altitude, and spectroscopic studies provide valuable information about the spicules through changes in the profile of the spectral lines. Doppler shift in these lines determines the velocity in the line of sight (Sekse, Rouppe van der Voort, and De Pontieu, 2012 and Sekse et al., 2013, Bose et al., 2019 and Bose et al., 2021a,b, Kuridze et al., 2015) and its changes with time and altitude from the surface of the Sun (Tavabi, Koutchmy, and Golub, 2015). From spectral line broadening, it is possible to measure non-thermal rotational velocities that lead to the indirect observation of torsional Alfvén waves. Spicules can heat the corona both by expelling hot plasma and by transferring energy by magneto-hydrodynamic waves (Rouppe van der Voort et al., 2009; Zeighami et al., 2016). So far, two types of spicules have been identified (De Pontieu et al., 2007a,b; Pereira, De Pontieu, and Carlsson, 2012). Type I spicules have a typical lifetime of 150 – 400 seconds with upward, and downward motions on parabolic paths with a maximum upward speed of 15 – 40 km s−1, while type II spicules have only upward motions with a lifetime up to 50 – 150 seconds, and a speed of 30 – 110 km s−1 (see Figure 4 in Pereira, De Pontieu, and Carlsson, 2012). Rouppe van der Voort et al. (2009) studied statistical properties of the disk counterparts of type II spicules from observations of Rapid Blue-shifted Excursions (RBEs) and found that the appearance, lifetimes, longitudinal and transverse velocities, and occurrence rate of these rapid blue excursions on the disk are very similar to those of the type II spicules at the limb. Pereira, De Pontieu, and Carlsson (2012) investigated the type I spicules and found relations between their properties with a magnetoacoustic shock wave driver and with dynamic fibrils as their disk counterpart. They also found that the properties of type II spicules are consistent with the properties of RBEs, confirming the hypothesis that RBEs are their disk counterparts. Type II spicules are mainly present in both the quiet Sun and coronal holes but in the active regions, type I spicules are predominant (Pereira, De Pontieu, and Carlsson, 2012). A recent study of observations from the Hinode and IRIS satellites showed that type II spicules are heated up to TR and coronal hole temperatures (Pereira et al., 2014; De Pontieu et al., 2011). The disk counterparts of type II spicules have been identified through the asymmetry of short-lived chromospheric spectral lines and are inferred from the rotation of blue or redshifts (Langangen et al., 2008; Rouppe van der Voort et al., 2009; Kuridze et al., 2015; Bose et al., 2019). Image processing and spectral methods are used to analyze the motion of structures such as spicules and solar jets. In the image processing method, displacements and intensity changes in images are measured (Tavabi et al., 2015a). In the second method, periodic patterns in Doppler shifts are examined by spectrometers (Tavabi, Koutchmy, and Golub, 2015, Tavabi, 2018; Tavabi and Koutchmy, 2019, and Zeighami, Tavabi, and Amirkhanlou, 2020).

In this study, we investigate solar off-limb EEs based on IRIS observations and then analyze the line spectra on 17 August 2014 and 19 February 2014. IRIS provides a tool for analyzing the thermal structure of the solar atmosphere in the UV wavelengths with high spatial and temporal resolution (De Pontieu et al., 2014b).

## 2. Observations

In this study, we analyzed two level 2 datasets from the IRIS database. IRIS spacecraft contained a combination of telescope and spectrograph (De Pontieu et al., 2014b). IRIS obtains bservations of the solar transition region with a high spatial and spectral resolution to investigate the physical processes of fine structures (Chen et al., 2019, Tavabi and Koutchmy, 2019, Zeighami, Tavabi, and Amirkhanlou, 2020, Bose et al., 2021a). The spectral dispersion in the far and near-ltraviolet are about 0.025 Å and 0.053 Å, respectively. The first data with OBSID 380011404 is a very large sit-and-stare observation of a limb filament target, taken on 17 August 2014 from 10:06:13 UT to 13:59:48 UT. The slit-jaw images (SJIs) field of view is 175" × 167" with cadence 32 s in both the 2796 Å and 1400 Å filters. The slit is centered at X = 827" and Y =−465" in heliocentric coordinates with a roll angle of −45 degrees. The cadence of the spectral observation is 16 s. The spatial pixel size is about 0.167" for both the spectral and SJIs. The second dataset OBSID 3800258253, is a large sit-and-stare observation of a limb spicules target, taken on 19 February 2014, from 16:27:46 UT to 17:35:45 UT. The SJIs' field of view is 119" × 119" , with cadences 23 s and 19 s in the 1330 Å and 1400 Å filters, respectively. The slit is centered at X = 6" and Y =−980" in the heliocentric coordinate with no roll angle. The cadence of this data is 9 s. The top panels of Figure 1 show the SJIs in 2796 Å and 1400 Å taken from IRIS, on 17 August 2014, and the bottom panels show the SJIs in 1330 Å and 1400 Å taken from IRIS, on 19 February 2014. The vertical lines in the panels refer to the location of the slit. It is noted that the effect of solar rotation for this, about 3-hour dataset, can be neglected.

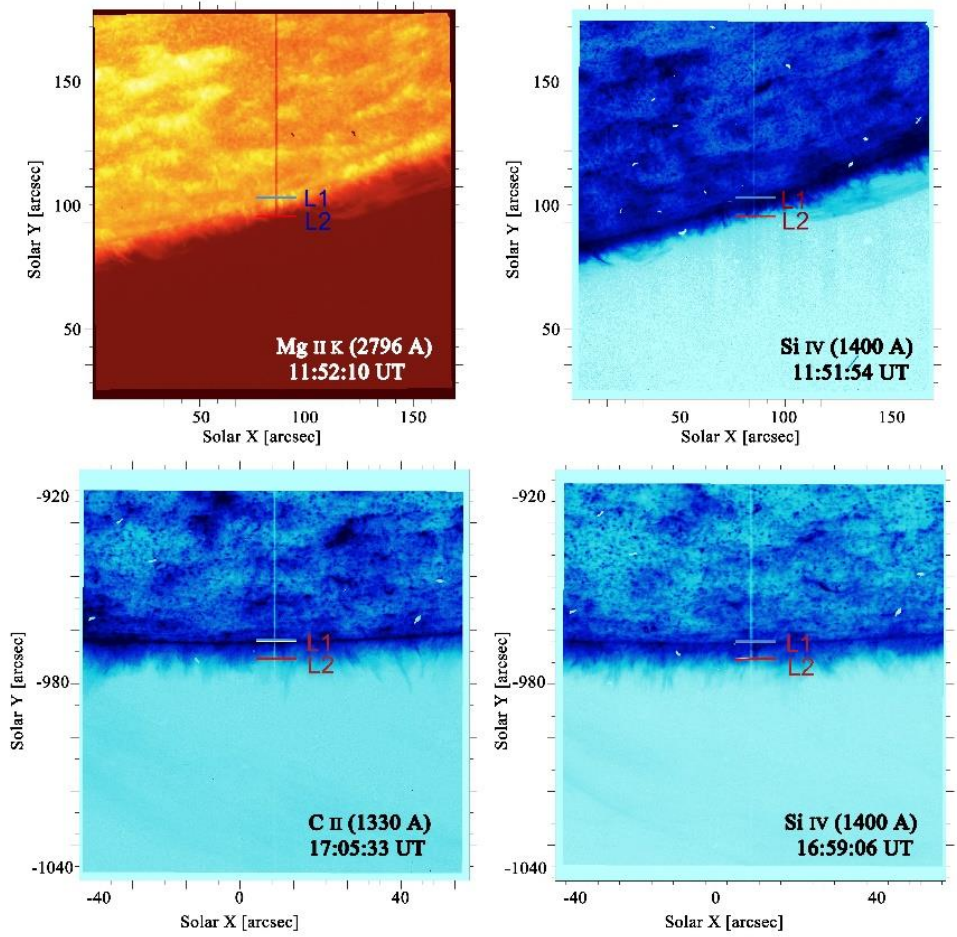

**Figure 1** Top panels show the SJIs in 2796 Å and 1400 Å filters taken from IRIS, recorded on 17 August 2014. The bottom panels show the SJIs in 1330 Å and 1400 Å taken from IRIS, recorded on 19 February 2014. The vertical lines in the panels refer to the location of the spectrograph slit, and two short green lines refer to altitudes along which time slices are calculated.

## 3. Data Analysis

For analyzing spectral images, we first apply an unsharp mask method for all images, and then select a time interval in which Doppler shifts can be seen with significant fluctuations (red and blue shifts) with respect to the central wavelength. The ranges for the two datasets are about 106 minutes and 68 minutes, respectively. Blue- and redshift indicate plasma motion towards and away from the observer, respectively. We obtain intensity profiles from the spectral images along two altitudes, i.e., at the solar limb and $y = 5\_\_$ away from the solar limb. In this way, time slice images of the profiles can be made. Some profiles are enhanced at both the blue and red wings, showing two peaks with comparable magnitudes, while other profiles are enhanced only at the blue wing or red wing. Chen et al. (2019) obtained similar profiles. The left panels of Figure 2 represent examples of Si IV 1403 Å and Mg II k 2796 Å spectra obtained on 17 August 2014. The two white lines refer to altitudes along which time slices are calculated, and the right panels show the temporal evolution of Si IV 1403 Å and Mg II k 2796 Å along two latitudes indicated by two white lines in the left panels. The left panels of Figure 3 represent examples of Si IV and Mg II k spectra obtained on 19 February 2014. The two white lines marked as L1 and L2 refer to altitudes along which time slices are calculated, and the right panels show the temporal evolution of Si IV 1403 Å and Mg II k 2796 Å along two latitudes indicated by two white lines on the left panels. Figure 4 shows the examples of the EEs with non-Gaussian profiles in Si IV 1403 Å line. Panels a – c display EEs from the first and panels d – f from the second dataset. Blue, both blue and red, and redshift are seen from left to right, respectively. Tavabi, Koutchmy, and Golub (2015) studied short-duration but complex jets related to a limb-event brightening (LEB) or limb EE directly above the south polar limb. At the Si IV emission temperature, they observed a pair of rapidly moving separated parts looking like two split branches (emission lines exhibit Doppler shifts toward red and blue, as one would expect from a bidirectional jet (Innes et al., 1997) at TR temperatures, with a narrow 400 km space between them in the radial direction and another part nearby rapidly moving in the opposite direction. Bidirectional jets in such events are accelerated in opposite directions from the reconnection sites. The first approximation null-point reconnection geometry could imply the involvement of magnetic loops, separatrix surfaces, and spine for which outflow jets are expected to have a multidirectional structure, which indeed has been observed in the form of simultaneous blue and red Doppler shifts in the limb EE. The bidirectional plasma jets ejected from a small reconnection site are interpreted to be the result of chromosphere or TR small loop–loop interactions that form in serpentine emerging flux that lead to reconnection. Si IV line is optically thin, except for events such as flares, but the Mg II lines are optically thick lines with a strong central reversal in the line cores (see Figure 5). Leenaarts et al. (2013a,b) studied the general formation mechanism of this line. Bose et al. (2019, 2021b) showed that the position of the maximum intensity marked as k2 is significantly affected by opacity effects in the solar chromosphere and the Doppler-shift of the line center named k3. The k3 line minimum is used as an actual representative of the mass flow. For eliminating cosmic rays and finding the Doppler velocity, we subtract the average of all spatial and temporal profiles from each profile per pixel (Tavabi and Koutchmy, 2019). Peter et al. (2014) studied hot explosions in the cool atmosphere of the Sun and demonstrated C II and the Mg II lines show a structure similar to Si IV farther away from the selfabsorption center. They found that it is reasonable to assume that the plasma ejected in the explosion covers temperatures from 6000 K to 100 000 K. This supports the conclusion that they observe a multi-thermal reconnection outflow. In the bombs, the Mg II lines show profiles that are very similar to those of Si IV. Therefore it might be that the outer parts (in the wing) of Mg II are an indication of the reconnection outflow, indicating a strong flow with a magnitude similar to what is also seen in Si IV in the cooler plasma where Mg II originates. The maximum intensity of the line spectrum is not the main true line of sight (LOS) velocity; however, it shows certainly true evidence of averaged Doppler velocities and more trustable. They showed an average spectrum for the limb above the quiet Sun region with the C II, Si IV, and Mg II lines. The line profiles generally conform to a single Gaussian, where under optically thin conditions. The lines of Mg II show self-absorption in the center due to large opacity. According to this suggestion and because all absorption lines are shifted, they are satisfied that the emission lines seen in the C II, Si IV, and MgII are primarily due to chromospheric (and TR) material directly above the EEs that is lifted in the flux-emergence process and cause the Doppler shifts. After determining the Doppler velocity at every time slice in Figures 2 and 3 and for exact analysis of these EEs, we plot separately the Doppler velocity versus time. Figure 5, from top to bottom, shows the line profiles of Mg II k spectra plotted as a function of velocity recorded on 17 August and 19 February 2014, respectively. Figure 6 shows IRIS Mg II k 2796 Å line spectra (time-averaged) plotted versus wavelength, and Y position (heliocentric) recorded on 19 February 2014. Figure 7, from top to bottom, illustrates the temporal evolution of Mg II k 2796 Å and Si IV 1400 Å spectra for the two altitudes, which were recorded on 17 August and 19 February 2014. The red circles indicate the position of Doppler velocity on the time slices of Figures 1 and 2. As Figure 7 shows, it can be seen that EEs have oscillatory behavior. For investigating the correlation between limb and off limb intensity oscillations, the two profiles are obtained. Figure 8 indicates the temporal evolution of spectra intensities on 19 February 2014 from Mg II 2796 Å line intensity for two signals with a height difference Y = 5" . The blue and red colors refer to the first signal near the limb and the second signal at a distance Y = 5" , respectively. Figure 9 indicates the temporal evolution of Doppler velocity obtained on 19 February 2014 from Mg II k 2796 Å line intensity for two signals with height difference colour coded the same way as in Figure 8. As Figure 9 shows, there is a correlation between two signals at certain times. We calculated the phase velocity of the two signals using a technique based on cross-correlation (Zeighami, Tavabi, and Amirkhanlou, 2020). In this method, the time lag between the two signals is calculated, and then by considering the distance between two altitudes (Y = 4 or 4000 km), the phase velocity is calculated. The phase velocity is obtained at about 220 km s$^{-1}$.

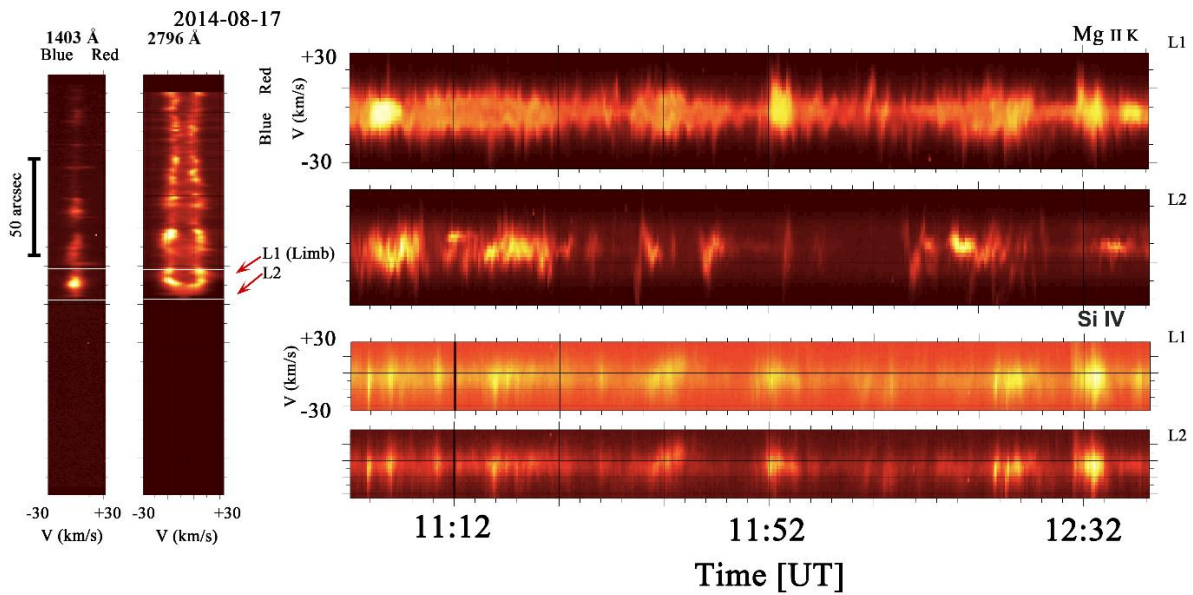

**Figure 2** Left panels represent examples of Si IV 1403 Å and Mg II k 2796 Å spectra observed on 17 August 2014. The two white lines named L1 and L2 refer to altitudes along which time slices are calculated. The right panels show the temporal evolution of Si IV 1403 Å and Mg II k 2796 Å along two altitudes indicated by two white lines in the left panels.

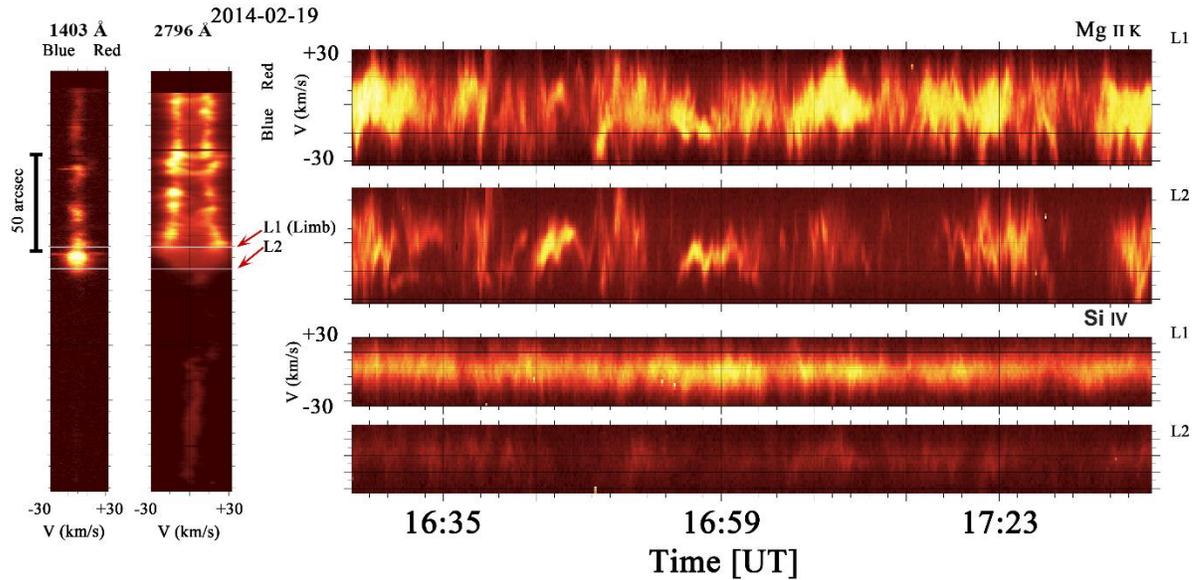

**Figure 3** Left panels represent examples of Si IV 1403 Å and Mg II k 2796 Å spectra recorded on 19 February 2014. The two white lines refer to altitudes along which time slices are calculated. The right panels show the temporal evolution of Si IV 1403 Å and Mg II k 2796 Å along the two altitudes indicated by two white lines in the left panels. The horizontal lines in the right panels indicate the rest wavelength position Mg II k 2796 Å and Si IV 1403 Å spectra.

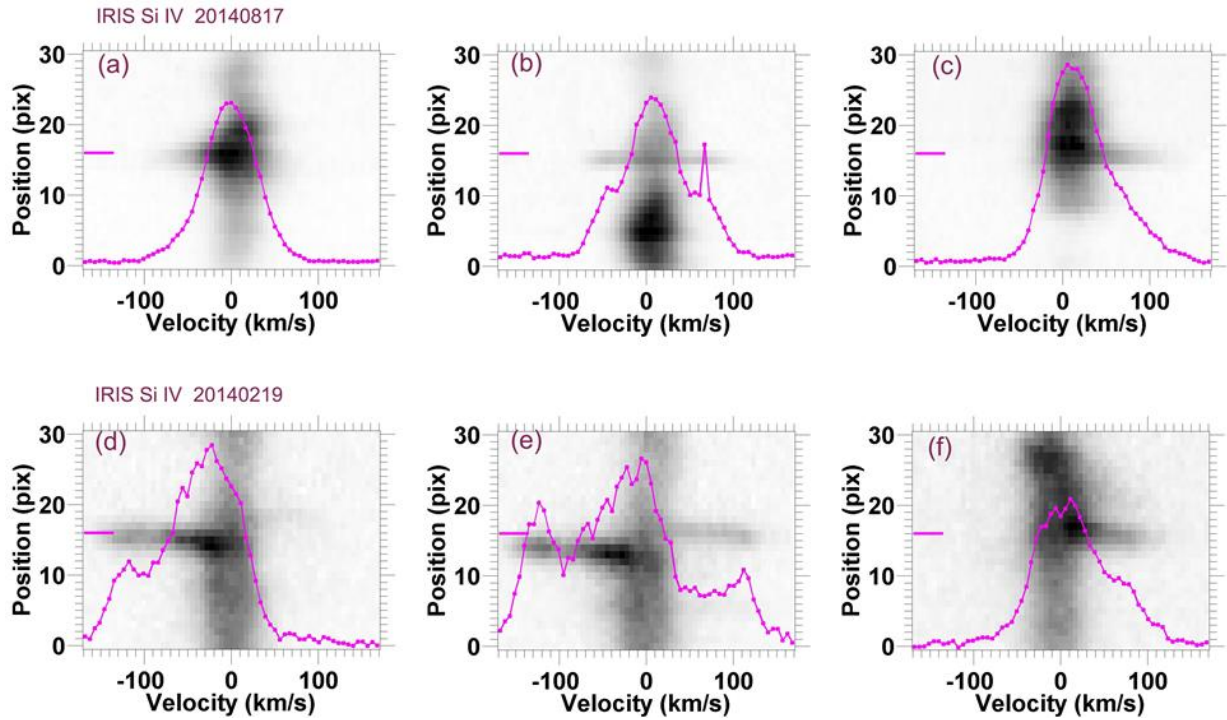

**Figure 4** Examples of the EEs with non-Gaussian profiles in Si IV 1403 Å line. Panels a—c show data from the first dataset on 17 August 2014, and panels d—f from the second dataset. Blue, blue and red, and edshift are shown from left to right, respectively

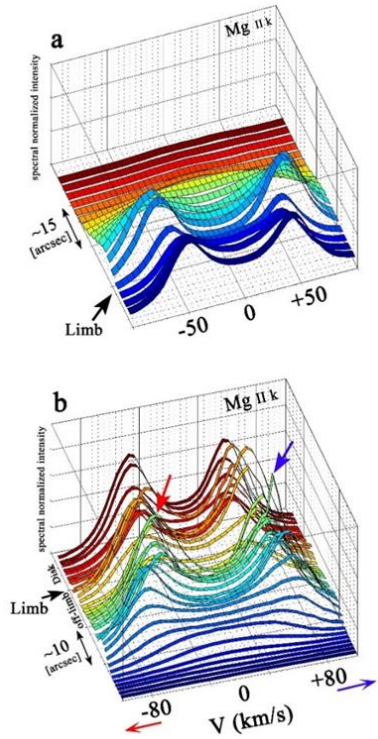

**Figure 5** Top and bottom panels are the   line profiles of Mg II k 2796 Å spectra plotted as a function of velocity recorded on 2014 August 17 and   February 19, respectively.

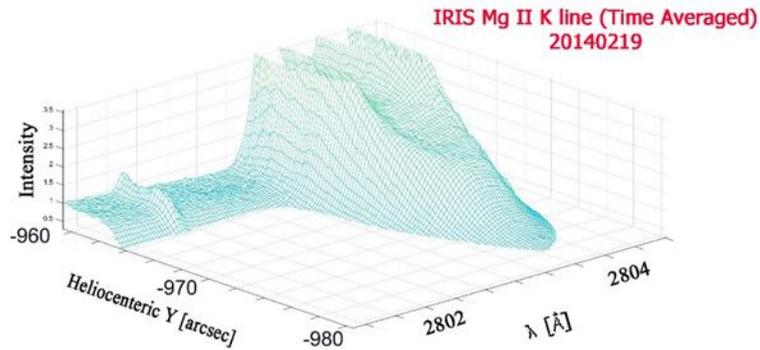

**Figure 6** IRIS Mg II k 2796 Å line spectra (time-averaged) are plotted versus wavelength, and Y position (heliocentric) recorded on 19 February 2014.

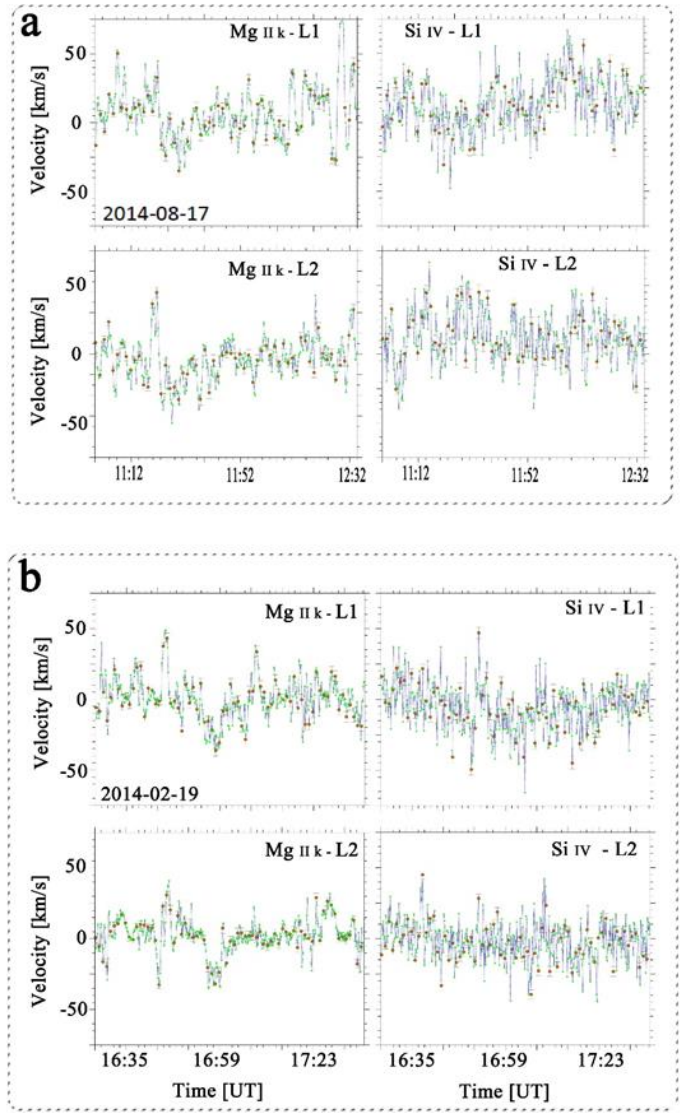

**Figure 7** (a) and (b) Temporal evolution of Doppler velocity obtained from images taken in Si IV 1403 Å and Mg II k 2796 Å spectra on 17 August and 19 February 2014, respectively. The red circles indicate the Doppler velocity along the time slices of Figures 2 and 3.

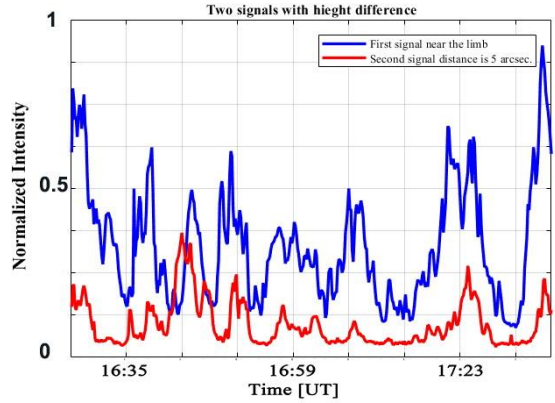

**Figure 8** Normalized intensity of two signals with height difference taken on 19 February 2014, the first and second signals are shown in blue and red colors, respectively.

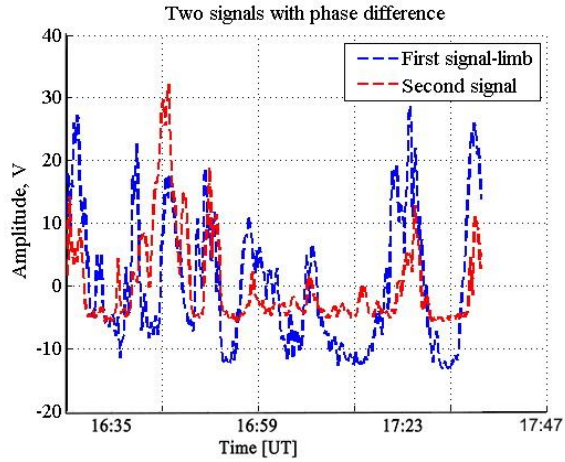

**Figure 9** Temporal evolution of Doppler velocity obtained on 19 February 2014 from Mg II k 2796 Å line intensity for two signals with height difference. The blue and red dashed lines refer to the first signal (near the limb) and the second signal (with a distance of Y = 5" ), respectively.

## 4. Conclusion and Discussion

In this study, using spectral images obtained from the IRIS Telescope by spectral analysis, we were able to detect Doppler shift fluctuations of EEs. By analyzing the spectral profiles for two altitudes across the slit, we determined the amplitude of the Doppler velocity for two datasets. Doppler velocity in the first dataset for the two altitudes was obtained from  about −50 km s$^{−1}$ to 50 km s$^{−1}$. De Pontieu et al. (2014b) studied the prevalence of small-scale twists in the solar chromosphere and transition region and found rapidly evolving twisting motions are apparent as short-lived, bright features in the blue and red wings around ±50 km s$^{-1}$ of the chromospheric Mg II h 2803 Å spectral line. We calculated the phase velocity of the oscillations using a technique based on cross-correlation. The phase velocity is obtained as about 220 km s$^{-1}$. As the spicules are almost in the vertical direction on the slit, the resulting velocities due to fluctuations along the line of sight can be considered the transverse velocities of the EEs. It should be noted that most component of the Doppler shifts in these observations related to the events shown in this research are in the transverse direction to local surface normal. The contribution of kink modes, which are parallel to the plane of the sky, cannot be detected by the observer because such Doppler velocities have no component along the line of sight. Photospheric convective motions are often the source of wave excitation in magnetic tubes (Zaqarashvili, 2003, Zaqarashvili et al., 2007). Therefore, the propagation of waves in the chromosphere can be traced through the dynamics of the EEs. In De Pontieu, Erdélyi, and James (2004), it is suggested that the high heat and density of materials inside the EEs, and, most importantly, their tilt increase the period of wave propagation inside them so that even waves with 5-minute periods can propagate. Magnetic tubes conduct three types of waves: kink waves, sausage waves, and rotating Alfvén waves. Some of these waves can cause Doppler displacement observations. Rotating Alfvén waves in thin tubes cause periodic non-thermal flattening of spectral lines but do not cause Doppler displacement fluctuations (Zaqarashvili, 2003). Sausage waves cause fluctuations in the intensity of the lines due to density changes, and if the axis of the tube is at an angle to the vertical line, the longitudinal velocity field of the sausage waves can cause Doppler displacement changes. However, the main contribution to Doppler displacement fluctuations is due to kink waves, which fluctuate transversely perpendicular to the axis of the tube (Zaqarashvili et al., 2007). In Pasachov, Jacobson, and Sterling (2009), the unsigned line of sight velocities of the spicules were determined to be slightly lower than 10 km s$^{-1}$. In De Pontieu et al. (2007b) and Pereira, De Pontieu, and Carlsson (2012), the velocities perpendicular to the main axis of the spicules were determined between 5 and 30 km s−1. For example, in De Pontieu et al. (2012), two types of transverse motions in spicules are identified: 1) Swaying motions of 15 to 20 km s$^{-1}$ and 2) Torsional motions in the range of 24 to 20 km s$^{-1}$. They reported that Alfvén waves with amplitudes of 10 to 25 km s$^{-1}$ with periods of 100 to 500 seconds could penetrate the chromosphere. When plasma flows in opposite directions from the source, the Doppler effect could be generated. In this case, strong enhancement in TR spectra is observed and has been known as bi-directional jets in which plasma moves along magnetic field lines (Curdt and Tian, 2011). Such Doppler velocities can be generated by any mechanisms that can produce bi-directional flows within the space of the pixel size. Explosive events are usually originated from magnetic reconnections (Hong et al., 2014), and the wing enhancement indicates bi-directional reconnection jets with speeds of 50 to 200 km s$^{-1}$. Rouppe van der Voort et al. (2015) analyzed short-lived asymmetries in chromospheric spectral lines based on IRIS and Swedish Solar Telescope (SST). They found clear signatures of rapid blue- or red-shifted excursions (RBEs or RREs) in Mg II h & k with asymmetries in spectral profiles and spectral signatures for RBEs and RREs in C II 1335 and 1336 Å and Si IV 1394 and 1403 Å spectral lines, that was interpreted as disk counterpart of type II spicules in upper chromosphere. In Tavabi and Koutchmy (2019), the authors investigated the properties of the spiculesmaterial off-limb above the 2.2 Mm heights using IRIS observations. They suggested that numerous small-scale jet-like spicules show rapid twisting and swaying motions evidenced by the large distortion and dispersion of the line profiles, including impressive periodic Doppler shifts, with an average swaying speed of order ±35 km s$^{-1}$ reaching a maximum value of 50 km s$^{-1}$ in the polar coronal hole region. They also identified for the first time waves with a short period of the order of 100 s and less and transverse amplitudes of the order of ± 20 to 30 km s$^{-1}$ with the definite signature of Alfvén waves. In Tei et al. (2020), by analyzing the Mg II spectrum, the speed of the spicules along the field of view was obtained from −25 to 25 km s$^{-1}$ so that the average asymptomatic speed of the spicules was from 2 to 10 km s$^{-1}$. According to the Doppler effect, the extended strong wings in the TR spectra indicate the existence of opposite-directional plasma flows in the source. Therefore, transition region EEs (TREEs) has also been known as "bi-directional jets". Chen et al. (2019) studied a statistic analysis at Si IV spectra to find the connection between TREEs and network jets. They found several types of EEs: TREEs with double peaks or enhancements in both wings located at either the footpoints of network jets, or transient compact brightenings, which are most likely due to magnetic reconnection; TREEs with enhancements only at the blue wing located above network jets along their propagation directions which clearly result from the superposition of the high-speed network jets on the TR background, TTEEs showing enhancement only at the red wing of the line generally located around the footpoints of network jets, likely due to the superposition of reconnection downflows on the background emission, and TREEs not corresponded to network jets associated with small-scale transient bright points in the 1330 Å SJIs, which could be characterized as quiet-Sun UV bursts.

Brueckner and Bartoe (1983), called as "turbulent events", the size varies from less than 1" , which is the resolution limit of the instrument, to 4" . The line profiles of TEs show a  strong, non-Gaussian enhancement of both the long-wavelength and short-wavelength wings (C IV, 1 × 105 K and also seen in C II 2 × 104 and N V 2 × 105 K). However, asymmetries in both directions have been observed at velocities ranging from ±50 to ±250 km s−1. Hansteen et al. (2014) investigated a similar phenomenon and named as unresolved fine structure (UFS); they were aligning the slit along the limb; IRIS also allows one to gather spectral data of the UFS. The spectrum shows large excursions as a function of position along the loop, implying large plasma velocities towards the red as well as towards the blue. They found extreme line profiles at the upper loop footpoint – the portion of the loop which meets the underlying atmosphere – during the entire 200 s lifetime of the UFS loop, with redward excursions of 70 – 80 km s$^{-1}$. This is two to three times the speed of sound in an 80,000 – 100,000 K plasma. They observed the spectral properties of several UFS loops and find that such high velocities often occur, though not always. This indicates that the UFS loops are locations of episodic and violent heating. About Alfvénic structure, the asymmetric (with a tilt angle) profiles on the red and blue wings of the events could be explained by a rotational ejection. Tavabi, Koutchmy, and Golub (2015) studied in detail the position of the EEs, and showed the very rapid blueand co-temporal redshifts in the C II and Mg II k raster Dopplergrams in the velocity space, that described in the literature as Alfvénic waves propagation (Brueckner and Bartoe, 1983; Martinez-Sykora et al., 2015; Sadeghi and Tavabi, 2022). Rompolt (1975) interpreted all shapes of spectral features (lines) from the expected rotation and expansion. He suggested  that the observed inclination to the direction of dispersion of some spectral features can be produced by rotation but not by expansion based on observational evidence. De Pontieu et al. (2012) showed that the tilted-streak morphology indicates relative redshift on one side of a spicule, blueshift on the other. This reversal in transverse motion is the signature of torsional spicule motion. In addition, the substantial offsets of the tilted streaks from the nominal line center can be understood as the upper position of swaying motion and a projection of field-aligned flow onto the line of sight. Tavabi, Koutchmy, and Ajabshirzadeh (2011) used the time slice Hinode Ca II H line with unprecedented spatial resolution and demonstrated that this spectrum tilt angle is related to the rotational motion such as Alfvénic torsional behavior. In conclusion, according to the periodic red and blue enhancements in EEs, we suggested that the fluctuations in the EEs with one side enhancement indicate the swaying motions of spicules over their axes, and those EEs observed on both wings indicate the rotational motions of spicules. The swaying and rotating motions are responsible for kink and torsional waves, respectively.

**Acknowledgments** We warmly acknowledge the work of our referee, who provided an extended detailed report and added many interesting suggestions and requests for greatly improving the paper. The authors gratefully acknowledge the use of data from the IRIS databases. IRIS is a NASA small explorer mission developed and operated by LMSAL with mission operations executed at NASA Ames Research center and major contributions to downlink communications funded by ESA and the Norwegian Space Centre.

**Declarations**

Disclosure of Potential Conflicts of Interest The authors declare that they have no conflicts of interest.